\documentclass[11pt]{article}
\pdfoutput=1
\usepackage{amsmath}
\usepackage{amssymb}
\usepackage{graphicx,bbm,mathrsfs}
\usepackage{nicefrac}
\usepackage{slashed}
\usepackage{bbm}
\usepackage{geometry}
\usepackage{empheq}
\usepackage{stackrel}
\usepackage{ulem}
\usepackage{jheppub}
\usepackage{setspace}
\usepackage{relsize}
\usepackage{empheq}
\usepackage{pifont}% http://ctan.org/pkg/pifont
\usepackage{mathtools}
\usepackage{enumitem}
\usepackage{dcolumn}   % needed for some tables
\usepackage{bm}        % for math
\usepackage{graphicx,mathrsfs}
\usepackage{nicefrac}
\usepackage{multirow}
\usepackage{color}
\usepackage{mathtools}
\usepackage{makecell}
\usepackage{environ} 
\usepackage{lipsum} 
\usepackage{wasysym}
\usepackage{hhline,colortbl}  
\usepackage{tikz}
\usepackage{graphicx}
\usepackage{tensor}
\usepackage{dsfont}
\usepackage{simpler-wick}
\usepackage{hyperref}

\usepackage{booktabs}
\usepackage{mdframed}
\usepackage{xcolor}

 \NewEnviron{Smaller11}{
           \scalebox{1.1}{$\BODY$} 
 } 
 \NewEnviron{Smaller08}{
           \scalebox{0.8}{$\BODY$} 
 }

\newcommand{\be}{\begin{equation}}
\newcommand{\ee}{\end{equation}}
\newcommand{\bea}{\begin{eqnarray}}
\newcommand{\eea}{\end{eqnarray}}

\newcommand{\R}{R}

\newcommand{\mLD}{\mu}

\newcommand{\nLD}{\textrm{LD}}
\newcommand{\nGL}{\textrm{GL}}

\renewcommand\labelenumi{(\roman{enumi})}
\renewcommand\theenumi\labelenumi

\newcommand{\nn}{\nonumber}

\usepackage{titlesec}

\titleformat*{\section}{\Large\bfseries}
\titleformat*{\subsection}{\large\bfseries}
\titleformat*{\subsubsection}{\large\bfseries}
\titleformat*{\paragraph}{\large\bfseries}
\titleformat*{\subparagraph}{\large\bfseries}

\makeatletter
\newcommand*{\prodsym}{%
  \DOTSB
  \mathop{
    \mathchoice
      {\rlap{\kern.3em\rotatebox[origin=c]{-90}{}}{\prod}}
      {\vcenter{\rlap{\kern.2em\rotatebox[origin=c]{-90}{}}}{\prod}}
      {\sum}{\sum}
  }\slimits@
}
\makeatother

\makeatletter
\DeclareFontFamily{OMX}{MnSymbolE}{}
\DeclareSymbolFont{MnLargeSymbols}{OMX}{MnSymbolE}{m}{n}
\SetSymbolFont{MnLargeSymbols}{bold}{OMX}{MnSymbolE}{b}{n}
\DeclareFontShape{OMX}{MnSymbolE}{m}{n}{
    <-6>  MnSymbolE5
   <6-7>  MnSymbolE6
   <7-8>  MnSymbolE7
   <8-9>  MnSymbolE8
   <9-10> MnSymbolE9
  <10-12> MnSymbolE10
  <12->   MnSymbolE12
}{}
\DeclareFontShape{OMX}{MnSymbolE}{b}{n}{
    <-6>  MnSymbolE-Bold5
   <6-7>  MnSymbolE-Bold6
   <7-8>  MnSymbolE-Bold7
   <8-9>  MnSymbolE-Bold8
   <9-10> MnSymbolE-Bold9
  <10-12> MnSymbolE-Bold10
  <12->   MnSymbolE-Bold12
}{}

\let\llangle\@undefined
\let\rrangle\@undefined
\DeclareMathDelimiter{\llangle}{\mathopen}%
                     {MnLargeSymbols}{'164}{MnLargeSymbols}{'164}
\DeclareMathDelimiter{\rrangle}{\mathclose}%
                     {MnLargeSymbols}{'171}{MnLargeSymbols}{'171}
\makeatother

\begin{document}

\vspace*{4mm}

\thispagestyle{empty}

\begin{center}

%  {\LARGE
% \sc
\begin{minipage}{20cm}
\begin{center}
\hspace{-5cm }
\huge
\sc
 Stable Black Strings 
\\  \hspace{-5cm }    from   Warped   Backgrounds
\end{center}
\end{minipage}
\\[30mm]

\renewcommand{\thefootnote}{\fnsymbol{footnote}}

{\large  
Sylvain~Fichet$^{\, a}$ \footnote{sylvain.fichet@gmail.com}\,, 
Eugenio~Meg\'{\i}as$^{\, b}$ \footnote{emegias@ugr.es}\,
Mariano~Quir\'os$^{\, c}$ \footnote{quiros@ifae.es}\,, 
Geovanna~Yamanaki$^{\, a}$ \footnote{elvira.yamanaki@ufabc.edu.br}\,, 
}\\[12mm]
\end{center} 
\noindent

${}^a\!$ 
\textit{CCNH, Universidade Federal do ABC,} 
\textit{Santo Andre, 09210-580 SP, Brazil}

${}^b\!$ 
\textit{Departamento de F\'{\i}sica At\'omica, Molecular y Nuclear and} \\
\indent \; \textit{Instituto Carlos I de F\'{\i}sica Te\'orica y Computacional,} \\
\indent \; \textit{Universidad de Granada, Avenida de Fuente Nueva s/n, 18071 Granada, Spain}

${}^c\!$  
\textit{Institut de F\'{\i}sica d'Altes Energies (IFAE) and} \\
\indent \; \textit{The Barcelona Institute of  Science and Technology (BIST),} \\
\indent \; \textit{Campus UAB, 08193 Bellaterra, Barcelona, Spain}

\addtocounter{footnote}{-4}

\vspace*{10mm}
 
\begin{center}
{  \bf  Abstract }
\end{center}
\begin{minipage}{15cm}
\setstretch{0.95}
% \small

We show that spacetime curvature alone can classically stabilize black strings.
Working within a consistent five-dimensional dilaton–gravity system with a flat brane, we find that sufficiently large black strings are classically stable when they extend from the brane to a timelike boundary, which may be either regular or conformal.
Black strings are also classically stable in the critical case of the linear dilaton spacetime.
In some of the curved backgrounds considered, black strings are stable despite having infinite horizon area.

    \vspace{0.5cm}
\end{minipage}

\newpage
\setcounter{tocdepth}{2}

\tableofcontents  

\vspace{1cm}
\hrule
\vspace{1cm}

\section{Introduction \label{se:Intro}}

Black strings  are vacuum solutions of the Einstein equations that can be viewed as higher-dimensional generalizations of black holes. For example, the simplest black string in a $D$-dimensional spacetime has cylindrical topology, $S^{D-2} \times \mathbb{R} $ \cite{Emparan:2008eg}. Such solutions exist in both asymptotically flat  and Anti-de Sitter (AdS) spacetimes, and have by now been extensively studied in the literature, 
see e.g. \cite{Gregory_1994,GregoryLaflamme1993,Gregory:1994EXTremal,Gregory:2000blackstring,Gregory:2011GLInstability,lehner2011finalstategregorylaflammeinstability,horowitz1992darkstringtheoryblack,Horowitz_1997,Hirayama,kang2002stabilityblackstringsbranes,BRIHAYE2008264,Dhumuntarao:2022,Figueras_2023,Harmark_2007,Kudoh_2006,Emparan_2022,Fernandes_Silva_2018,collingbourne_2022,Collingbourne_2021,Reall_2001,Konoplya_2008,Konoplya_2011,Aharony_2004, Horowitz:1991cd}.

In both flat and AdS spacetimes, black strings are classically unstable, a phenomenon known as the Gregory-Laflamme (GL) instability \cite{GregoryLaflamme1993,Gregory:2000blackstring}.
In flat space, the non-linear evolu\-tion of the GL instability 
most likely leads  to fragmentation of the black string into a sequence of  Schwarzschild black holes \cite{GregoryLaflamme1993,Gregory_1994}, possibly connected by smaller strings \cite{lehner2011finalstategregorylaflammeinstability}. In AdS, a similar fragmentation process occurs, with the difference that a droplet-shaped horizon anchored to the AdS boundary also remains \cite{Chamblin1999, Giddings:2000mu, Emparan_2022}.
The above features persist  in the presence of a flat $(D-2)$-brane perpendicular to the string.\,\footnote{From the low-energy viewpoint, a brane is simply an infinitely thin hypersurface living in the higher dimensional spacetime and on which operators and degrees of freedom can be localized \cite{Csaki:2004ay, Sundrum:1998sj,Sundrum:1998ns}. For the bulk fluctuations living on one side of the brane, the other side is irrelevant and can be ignored --- it is thus referred to as ``end-of-the-world" (EOW) brane. } 
The solutions exist analogously, with the black string ending on the brane, and  the GL analysis proceeds in the same way. 
% In the AdS case, a black droplet  anchored  to the brane remains, with  proper width of order  ${\cal O}(R_{\rm AdS})$\,\cite{Chamblin1999, Giddings:2000mu}. 

Is it possible  to obtain a  black string  that is classically stable? 
One possibility is to compactify  the spatial direction along which the string extends.  For example, the  black string may  be wrapped around a $S_1$, or it may stretch  across a compact interval \cite{Gregory:2000blackstring}. Here we follow a different path. 

% However, to keep the inter-brane distance finite, some stabilization mechanism is needed. This typically requires the introduction of a scalar field with appropriate potential, as described by e.g. the Goldberger-Wise model {\cite{Goldberger_1999}}.  Here we do not follow this path, as we will show that the presence of two branes is unnecessary to stabilize the black string.

In this work, we demonstrate that the curvature of spacetime itself can stabilize  black strings. 
 As we will show,  although in certain cases the warping of spacetime does have an effect of compactification, the  stabilization phenomenon  is actually more general. 

Concretely, we consider a scalar-gravity system in the presence of  a single flat brane. Consistent solutions to the field equations give rise to warped geometries that depart from AdS. 
In particular, the scalar field vacuum expectation value (vev) typically diverges at some point in the bulk, such that vacuum energy blows up and a curvature singularity appears (see e.g.~\cite{Gubser:2000nd,Gursoy:2007er,Cabrer:2009we, vonGersdorff:2010ht,Chaffey:2023xmz}). Several criteria have been proposed to assess  when such a curvature singularity is admissible (or ``good'') {\cite{Gubser:2000nd,Cabrer:2009we}}.  Here, as an aside, we present a new criterion based solely on the spectrum of gravitational fluctuations.

We focus on  a specific class of scalar potential that is well motivated by  string UV completions and dimensional compactifications (see e.g. \cite{Gursoy:2007cb}). For this reason we refer to the scalar field as the \textit{dilaton}.
The resulting class of spacetimes, denoted by $\mathcal{M}_{\nu}$ with $\nu$ a real parameter,  
includes AdS$_D$ and, importantly,  the linear dilaton spacetime LD$_D$ \cite{Fichet:2023xbu,Fichet:2023dju}. The latter arises from string-theoretical constructions: non-critical string theory \cite{Tong_lecture}, little string holography \cite{Aharony:1999ti} and its recent avatars \cite{Giveon:2017nie, Chang:2023kkq}.

In addition to the possible presence of a curvature singularity, the $\mathcal{M}_{\nu}$ spacetimes may exhibit a regular boundary or a conformal boundary. 
We  investigate how these  fundamental features of the background affect the stability of the black string.

One might wonder why to study black strings and their stability in the first place. 
Broadly, the study of black objects is a way to gain insights  to better understand gravity. 
For example, we will contrast our results with the correlated-stability conjecture (CSC) \cite{Gubser:2000ec,Gubser:2000mm}, that connects thermodynamics to classical stability. 
  Furthermore,  some of the  spacetimes we are considering are holographic. In this context, studying the black string is relevant to understand the properties of black holes in  the presence of a strongly coupled sector. Such a study is also of phenomenological relevance,  as some of the setups we consider yield realistic cosmological braneworld models in AdS (see e.g. \cite{Randall:1999vf, Shiromizu:1999wj,Binetruy:1999hy,Hebecker:2001nv,Langlois:2002ke,Langlois:2003zb}) and beyond \cite{Fichet:2022ixi, Fichet:2022xol, Fichet:2026nct}.

Our study is structured as follows. In section~\ref{se:BlackString} we present the consistent solving of the dilaton-gravity background and discuss the basic properties of the ${\cal M}_\nu$ spacetimes. 
We introduce the admissibility criterion for the singularity. 
We then present the black string solution. 
Section~\ref{se:GL_review} reviews the basics of the GL instability computation in asymptotically-flat space. 
In section~\ref{eq:fluctuations}, focussing on $D=5$, we introduce the
spacetime fluctuations and their gauge fixing, and systematically compute the spectrum of gravitational fluctuations. 
In section~\ref{se:stability_results} we compute  the GL instability using the spectrum. We show that in a subset of the ${\cal M}_\nu$ spacetimes, the black string is classically stable. We also discuss the correlations between classical stability and other aspects of the background.  Section~\ref{se:Conclu} summarizes these results. 

\subsubsection*{Conventions}

We use the Misner-Thorne-Wheeler conventions \cite{Misner:1973prb}, with mostly-plus spacetime signature ${\rm sgn}(g_{MN})=(-,+,+,...,+)$. The metric determinant is denoted $\sqrt{g} \equiv \sqrt{|g_{MN}|}$. In most cases, we study spacetimes with $D=d+1$ dimensions. Accordingly, $D$-dimensional coordinates are labeled by capital Latin indices ($M,N,...$), and $d$-dimensional coordinates by Greek indices ($\mu,\nu,...$).

\section{Dilatonic Spacetimes and the Black String \label{se:BlackString}}

%http://arxiv.org/abs/2406.02899% 
%http://arxiv.org/abs/2311.14233%
%http://arxiv.org/abs/2309.02489%

\subsection{Introducing Dilaton Gravity}
We consider a $D$-dimensional spacetime and write the general dilaton-gravity bulk action
\begin{equation}
\begin{split}
& \mathcal{S}{[g_{MN},\phi]}=
\frac{1}{2}\int d^{D}x \,\sqrt{g}\,
\Bigl(M_{D}^{D-2}\ ^{(D)}R
\;-\;(\partial_{M}\phi)^{2}
\;-\;2V(\phi)
\Big)\,,
\end{split}
\label{Acao-Scalar-Gravity}
\end{equation}
where $^{(D)}R$ is the curvature scalar in $D$ dimensions, $M_{D}$ is the fundamental $D$-dimensional Planck scale, $\phi$ is the dilaton field and $V(\phi)$ the scalar potential.
We introduce the reduced quantities $\bar{V}(\bar{\phi})$ and $\bar{\phi}\,,$

{\begin{equation}
\bar{V}(\bar{\phi}) = \frac{V(\bar{\phi})}{(D-2)M^{D-2}_{D}}\ \ \ {\rm with} \ \ \ \bar{\phi}=\frac{\phi}{\sqrt{D-2}M^{\frac{D}{2}-1}_{D}}\,,
    \label{Pot}
\end{equation}}
where $\bar{V}(\bar{\phi})$ has mass dimension 2 and $\bar{\phi}$ is dimensionless.

\subsubsection{Physics on the Brane} We describe a flat ($D-2$)-brane embedded in the $D$-dimensional bulk with induced metric $\bar{g}_{\mu\nu}= g_{MN} - n_{M}n_{N}$, in which $n_{M}$ is the normal vector to the brane. The brane action is given as follows,
\begin{equation}
    \mathcal{S}_{brane} = \;-\;
\int d^{D-1}x \,\sqrt{\bar g}\,
\bigl(V_{b}(\phi)+\Lambda_{b}-M_{D}^{D-2}K\bigr)\,.
\end{equation}
The brane supports a tension $\Lambda_{b}$ and a localized potential $V_{b}(\phi)$, that stabilizes the dilaton to its vev, $\left\langle\phi_{b}\right\rangle = v_{b}$. In the action, $K$ is the extrinsic curvature scalar that appears in the Gibbons-Hawking-York {(GHY)} boundary term. Here it suffices to assume $V_{b}(v_{b})=0$.  

    The dynamics of a brane in the bulk can be described by the Gauss-Codazzi equations {\cite{Shiromizu_2000,Nastase2019}}, or equivalently by Israel's junction conditions \cite{Israel:1966rt}. These relate the extrinsic curvature quantities on each side of the brane, $K_{\mu\nu}^{+}$ and $K_{\mu\nu}^{-}$, to the brane localized energy-momentum tensor $S_{\mu\nu}$ see e.g. \cite{Nastase2019,Fichet:2023xbu},
\begin{equation}
    K_{\mu\nu}^{+} - K_{\mu\nu}^{-} = -\frac{1}{M_{D}^{D-1}}\left(S_{\mu\nu}-\frac{1}{D-2}\bar{g}_{\mu\nu}S \right) \,,
    \label{junction}
\end{equation}
where $S_{\mu\nu}=\tau_{\mu\nu}-\Lambda_{b}\bar{g}_{\mu\nu}$, with $\tau_{\mu\nu}$ the matter component localized on the brane.
When  both sides  simply are the mirror of
each other, the brane can equivalently be described as a wall beyond which spacetime ends, i.e.~an EOW brane.  Since this is a boundary, one must let the  metric be dynamical on the brane. This yields a Neumann-type boundary condition that corresponds precisely to \eqref{junction} with  zero $K_{\mu\nu}$  on the empty side of the brane; see e.g.~\cite{Barbosa:2024pyn}. 

\subsubsection{Warped Spacetime and Field Equations} Assuming a general warped metric ansatz, we split the coordinates as $x^{M}=\{x^{\mu},r\}$ with $x^{\mu}=\{\tau,x^{i}\}$ in $d=D-1$ dimensions,
\begin{equation}
    ds^{2} = g_{MN}dx^{M}dx^{N} = e^{-2A(r)}\eta_{\mu\nu}dx^{\mu}dx^{\nu} + e^{-2B(r)}dr^2\,,
    \label{Line-Element-Ansatz}
\end{equation}
where $A(r)$ and $B(r)$ are the warp factors.
The independent field equations obtained from the general ansatz are
\begin{eqnarray}
    A''(r) + A'(r)B'(r) -\bar{\phi}' \ ^2 &=& 0\,,
    \label{Field-Equations-01} \\
    A'(r)^2 + \frac{1}{D-1}\left(2\bar{V}(\bar{\phi})e^{-2B(r)} - \bar{\phi} ' \ ^2\right) &=& 0\,.
    \label{Field-Equations-02}
\end{eqnarray}
At the level of solutions, one extra field equation implies an algebraic constraint on the integration constants. A careful treatment of the solutions is presented in Ref.~\cite{Fichet:2023xbu}. Once   gauge redundancies and  boundary conditions are taken into account, a single combination remains, that is identified as the physical scale of the geometry.

% Neverthless, a combination of such constants can be fixed by requiring the vev $\left\langle\phi_{b}\right\rangle$ to be independent of the brane location, once the brane localized potential is also independent on the brane position.

\subsection{The ${\cal M}_{\nu}$ Spacetimes}

We consider a  family of $D$-dimensional dilaton--gravity spacetimes~\cite{Fichet:2023xbu,Fichet:2023dju}, denoted ${\cal M}_\nu$, defined by setting the  bulk potential~\eqref{Pot} to an exponential,
\begin{equation}
\bar{V}(\bar{\phi}) = -\frac{1}{2} \left( D-1-\nu ^2 \right) k^2 e^{2\nu\bar{\phi}}\,,
\label{eq:V_Mnu}
\end{equation}
where $k$ is a dimensionful constant. 
At that point, the $\nu$ parameter is a real parameter that we can take positive without loss of generality.

We place the flat brane at $r=r_{b}$, dividing $\mathcal{M}_{\nu}$ in two regions, which we denote $\mathcal{M}^{-}_{\nu}$ and $\mathcal{M}^{+}_{\nu}$:

\begin{equation}
    \mathcal{M}_{\nu}^{-} = \mathcal{M}_{\nu}\Big|_{r\in(0,r_{b}]}\,, \ \ \ \ \ \ \ \ \ \ \ \  \mathcal{M}_{\nu}^{+} = \mathcal{M}_{\nu}\Big|_{r\in[r_{b},\infty)}\,. \label{eq:Mpmdef}
\end{equation}

The $\cal M_{\nu}$ spacetimes  satisfy the field equations (\ref{Field-Equations-01})-(\ref{Field-Equations-02}) with potential \eqref{eq:V_Mnu}. In the absence of any black hole, the solutions are  \cite{Fichet:2023dju}
\begin{eqnarray}
    ds^2_{\nu,r_b} &=& \Big(\frac{r}{L}\Big)^{2}\eta_{\mu\nu}dx^{\mu}dx^{\nu} + \Big(\frac{r}{r_{b}}\Big)^{2 \nu^2}\frac{1}{(\eta r)^2}dr^2\,,
    \label{sol01}  \\
        \bar{\phi}(r) &=& \bar{v}_{b} - \nu \ln \left( \frac{r}{r_{b}} \right) \,,
    \label{sol02}
\end{eqnarray}
 where $L$ is an integration constant with dimension of length,
 %that can be fixed to any value,
 and  {$\eta \equiv k \, e^{\nu \bar{v}_{b}}$}  is the physical mass scale that emerges from the reduction of integration constants, and that appears in observable quantities measured on the brane.

Given this solution, the non-zero components of the Ricci tensor are
\begin{equation}
\R_{ii} \;=\;-\,R_{\tau\tau}
\;=\;
(\nu^2 - (D-1))\,
\left(\frac{\eta r_b}{L}\right)^{2}
\left(\frac{r}{r_b}\right)^{2(1-\nu^2)}\,, \qquad R_{rr} =\frac{(D-1)}{r^2}(\nu^2-1)\,,
\label{RicciT-Ddim}
\end{equation}
{and} the curvature scalar is

\begin{equation}
\ \ \ \ \ \ R \;=\;
(D-1)(2\nu^2 - D) \eta^2
\left(\frac{r_b}{r}\right)^{2\nu^2}.
\label{Ricci-Ddim}
\end{equation}

In any dimension, $\nu=0$ corresponds to anti-de Sitter spacetime AdS$_D$, and $\nu=1$  corresponds to the linear dilaton spacetime LD$_D$.

% Spacetimes exhibit different curvature behavior depending on the value of $\nu$. For example, in five dimensions, setting $\nu=1$ in the bulk potential yields $\rm AdS_{5}$ spacetime, while choosing $\nu=0$ corresponds to the linear dilaton spacetime $\rm LD$ in five dimensions.\\ \\

\subsubsection{Conformal Coordinates} 

Throughout the work, we often make use of conformal coordinates. In the $\nu\neq1$ case these are
\begin{equation}
    ds^{2}_{\nu,r_{b}} = \;\Bigl(\frac{r_b}{L}\Bigr)^{\frac{2\nu^2}{\nu^2-1}}\,
\bigl(\lvert\nu^2-1\rvert\,\eta\,z\bigr)^{\frac{2}{\nu^2-1}}
\Bigl(\eta_{\mu\nu}\,dx^\mu dx^\nu + dz^2\Bigr) \,,
\label{ConformalDs-01}
\end{equation}
where $z$ is the {conformal} coordinate for the extra dimension, defined as
\begin{equation}
    z \;=\; \frac{L}{\eta\,r_b\,\lvert\nu^2-1\rvert}
\Bigl(\frac{r}{r_b}\Bigr)^{\nu^2-1}
\,,\qquad \nu\neq1,
\quad z\in\mathbb{R}_{+}\,.
\label{Conformalz-01}
\end{equation}
In such coordinates, the background spacetimes $\mathcal{M}_{\nu}$ are defined in terms of the  {value of the $\nu$ parameter} as
\be
{\cal M}_\nu^- =\begin{cases}
\mathcal{M}_{\nu}\big|_{z \ \in [z_{b}, \infty )} \quad {\rm if}\quad \nu <1
\\
\mathcal{M}_{\nu}\big|_{z \ \in (0,z_{b}]}  \quad~ {\rm if}\quad \nu >1
\end{cases},\quad \quad
{\cal M}_\nu^+ =\begin{cases}
\mathcal{M}_{\nu}\big|_{z \ \in (0,z_{b}]} \quad~ {\rm if}\quad \nu <1
\\
\mathcal{M}_{\nu}\big|_{z \ \in [z_{b}, \infty )}  \quad {\rm if}\quad \nu >1
\end{cases} \,, \label{eq:Mpmdefconf}
\ee
with $z_{b}$ the brane position in the new coordinate.

For  the special case of linear dilaton ($\nu=1$), we have
\begin{equation}
    z = \pm\frac{L}{r_{b}\eta}\log\frac{r}{L}\,, \ \ \ \ \ \ \ \ \ z\in\mathbb{R} \,,
    \label{Conformalz-02}
\end{equation}
and the line element is
\begin{equation}
    ds_{\nLD}^2 = e^{\pm 2\frac{r_{b}\eta}{L}z}(\eta_{\mu\nu}\,dx^\mu dx^\nu + dz^2) \,.
    \label{ConformalDs-02}
\end{equation}
The freedom of sign in \eqref{ConformalDs-02}  is reminiscent of a discrete  symmetry of the LD spacetime pointed out in \cite{Fichet:2023xbu}.  
Notice that $z\in \mathbb{R}$ for LD, instead of $\mathbb{R}_+$ for $\nu\neq 1$. 
 In the following we choose the $+$ convention, such that $ z = \frac{L}{r_{b}\eta}\log\frac{r}{L}$.  With this choice, the two regions of spacetime are 
 \be
 {\cal M}^-_1\equiv {\rm LD}^- = {\cal M}_\nu \Big|_{z\in (-\infty, z_b]}\,, \quad\quad
  {\cal M}^+_1\equiv {\rm LD}^+ = {\cal M}_\nu \Big|_{z\in [z_b,\infty)}\,.
 \ee

\subsubsection{A Graviton Criterion for Admissible Singularities}

The $\mathcal{M}^{-}_{\nu}$ spacetime presents a curvature singularity at $r=0$ for all $\nu>0$, while no singularity appears in $\mathcal{M}_{\nu}^{+}$ for any $\nu$. 
Such curvature singularities are usually considered to be physically admissible (or ``good'') using criteria on the boundedness of the bulk potential~\cite{Gubser:2000nd,Cabrer:2009we}. 
% The singularity is usually labeled as ``good'' \textcolor{red}{(provided that $\nu<\sqrt{d}$)} using criteria on the boundedness of the potential~\cite{Gubser:2000nd,Cabrer:2009we}. 
One can also argue that the singularity must be hidden by a black hole horizon \cite{Gubser:2000nd, Fichet:2023dju}. Following this, the singularity at $r=0$ in the ${\cal M}^-_\nu$ spacetime should be cloaked by a planar black hole. We derived the general  planar black hole solution in \cite{Fichet:2023dju,Barbosa:2024pyn}, with blackening factor $f(r) = 1-\left(\frac{r_h}{r}\right)^{d-\nu^2} $. This solution implies that the $\nu$ parameter is restricted to $\nu\in[0,\sqrt{d})$.

In the present work we {provide} a different criterion, solely based on the graviton spectrum and with no need of explicit black hole solution. Our argument is the following. 
If spacetime contains a curvature singularity, the graviton spectrum must have a zero mode. Otherwise,  gravity would decouple entirely at low energy.  This would imply that  nothing can forbid a signal to travel back and forth to the naked singularity of the spacetime background. The singularity being timelike, it causes causality violation:  a signal can travel to the singularity and come back from any different point in time.  The existence of a graviton zero mode is thus mandatory to ensure consistency of the theory.

The graviton spectrum in the ${\cal M}^-_\nu$ spacetimes will be computed in section \ref{eq:fluctuations}. We find that the graviton zero mode exists if
\be
\nu\in[0,\sqrt{d})\,.
\ee
This agrees exactly with \cite{Gubser:2000nd,Cabrer:2009we,Fichet:2023dju} --- which is fairly nontrivial given the different nature of  calculations.

\subsubsection{Boundaries \label{se:boundaries}}

We use the conformal coordinates \eqref{ConformalDs-01} to study the boundaries of ${\cal M}_\nu$. For $\nu<1$ we find that $\mathcal{M}_{\nu}^{+}$ has a conformal boundary at $z=0$. For $\nu>1$, $\mathcal{M}_{\nu}^{-}$ exhibits a regular boundary at $z=0$, that matches the  curvature singularity. 
In the  linear dilaton case ($\nu=1$), the boundaries are null.  The Penrose diagram is the same as Minkowski's, and the singularity lies at a spatial infinity of the causal diamond.

\subsection{The Black String Solution}

% We proceed by considering the special case of $D=5$. As will be clarified below, the study of $5$-dimensional black strings in the $\mathcal{M}_{\nu}$ background allows us to study the GL instability in spacetimes of particular interest, including $\rm AdS_{5}$ and $\rm{LD}_{5}$, which have previously shown to exhibit interesting features in an holographic and duality context \cite{Fichet:2023holographic}.

We finally specify $D=5$ for simplicity. We introduce spherical coordinates in the metric ~(\ref{Line-Element-Ansatz}),  with $ x^{\mu}=\{\tau,\rho,\theta,\varphi\}$, and allow for a blackening factor on the four-dimensional slice of the metric, 
\begin{equation}
    ds^{2}_{\rm BS}= \gamma^{\rm BS}_{MN}dx^M dx^N = e^{-2A(r)}\left(-U(\rho)d\tau^2+U(\rho)^{-1}d\rho^2 + \rho^2d\Omega_{2}^{2} \right) + e^{-2B(r)}dr^2\,.
    \label{Black_String_Line_Element}
\end{equation}
We obtain the solution   \be U(\rho) = 1 - \frac{\rho_{h}}{\rho}\,,\ee  in addition to the solutions \eqref{sol01} and \eqref{sol02} for $A(r)$, $B(r)$ and $\phi(r)$. This describes a  horizon at $\rho=\rho_h$, with topology of a cylinder.

The nonzero Ricci  components are
\begin{equation}
R_{ii} = -R_{\tau\tau} =
\frac{
  (4 - \nu^2)\, (\eta r)^2\,
  \bigl(\tfrac{r_b}{r}\bigr)^{2\nu^2}\,
}{
  L^2\,\rho
} (\rho_h - \rho)\,,
\end{equation}  
where $i \in\{ x^1, x^2, x^3 \}$ is the {3D spatial} index in cartesian coordinates, while $R_{rr}$ and the Ricci scalar remains the same as in \eqref{RicciT-Ddim} and~\eqref{Ricci-Ddim} for the black string. The Kretschmann scalar is given by:

\begin{equation}
    {\rm Riem}^2\equiv R_{\rho\lambda\sigma\gamma} R^{\rho\lambda\sigma\gamma} =  8\eta^4 (5-4\nu^2 + 2\nu^4)\Big(\frac{r_b}{r}\Big)^{4\nu^2} + \frac{12L^4 \rho_h^2  }{r^4\rho^6} \,.
        \label{kretchsmann}
\end{equation}

\subsubsection*{Discussion}

A simple qualitative analysis of $ {\rm Riem}^2$ provides us with valuable information. 
The black string and background contributions are cleanly split in \eqref{kretchsmann}. Based on this result, we can expect two qualitatively different regimes depending on $\nu$. We have ${\rm Riem}^2|_{\rm background}\ll {\rm Riem}^2|_{\rm string} $ at small $r$ for $\nu<1$ and at large $r$ for $\nu>1$.
These regions are respectively in ${\cal M}^-_{\nu<1}$ and  ${\cal M}^+_{\nu>1}$.  
Therefore, in these regions, the black string should behave approximately as in flat space. It should thus suffer some version of the GL unstability.~\footnote{A similar observation was made in \cite{Chamblin1999} for AdS black strings. } 
In contrast, black string stability might be expected in  ${\cal M}^-_{\nu>1}$ and  ${\cal M}^+_{\nu<1}$.

We will explicitly prove these facts in the next sections.

% The metric of \eqref{sol01} can be replaced by any other Ricci-flat metric, such as the Schwarzschild four-dimensional metric \cite{Gregory:2000blackstring}, preserving the previously solutions \eqref{sol01} and \eqref{sol02} derived for the case with no black hole. Thus, all $\mathcal{M}_{\nu}$ features presented so far—singularities, boundaries, and the presence of a \(3\)-brane—remain unchanged upon the inclusion of the black string. 

\section{Gregory-Laflamme Review}
\label{se:GL_review}

Unlike four-dimensional black holes, black strings and black $p$-branes are unstable under small classical perturbations~\cite{GregoryLaflamme1993}. This phenomenon is referred to as the Gregory–Laflamme (GL) instability. It is  established for a broad class of extended black objects  with and without translational invariance in asymptotically flat~\cite{Gubser_2002,Emparan_2003,Hovdebo_2006,Santos_2015,
Bantilan_2019,Figueras_2016,Emparam2002,Emparan:2008eg,Dhumuntarao:2022} and AdS spacetimes, see e.g.~\cite{Chamblin1999, Giddings:2000mu, Emparan_2022}.

The instability takes place at the level of linearized gravitational perturbations and can be identified by the presence of normalizable  modes that exponentially grow  in time. The instability is caused by perturbations whose wavelength along the extra dimension is large compared to the cross-section horizon area of the black string~\cite{Gregory_1994,GregoryLaflamme1993,Gregory:2000blackstring}.

\subsection{Instability in  Asymptotically Flat Spacetimes}

We review   the GL computation for the simplest black string. We consider
the 5D black string given by Schwarzschild$_4 \times \mathbb{R}$,
\begin{equation}
    ds^{2}_{\rm BS, flat} = \gamma^{\rm BS, flat}_{MN} dx^M dx^N =  -f(\rho)dt^{2}+ f(\rho)^{-1}d\rho^2 + \rho^{2}d\Omega^{2}_{2} \ + \ dy^{2}\,,
    \label{black_string_GL}
\end{equation}
where $y$ is the coordinate of the extra dimension and $f(\rho)$ is the Schwarzschild factor 
\begin{equation}
    f(\rho) = 1 - \frac{\rho_{h}}{\rho} \,.
\end{equation}
We introduce the perturbation $ g_{MN} = \gamma^{\rm BS, flat}_{MN}  + h_{MN}$. Invariance of the action under diffeomorphisms implies the gauge symmetry  $\tilde h_{MN}\to \tilde h_{MN} +\nabla_M \xi_N+\nabla_N \xi_M$, that can be used to remove five unphysical degrees of freedom. 
We choose  the transverse-traceless gauge $\nabla^M h_{MN}=0=h_M^M$.~\footnote{One could instead choose to set $h_{\mu y}=0=h_{yy}$ as in \cite{Fichet:2023dju}, leaving only the $h_{\mu\nu}$ component, but more terms remain in the equation of motion with this choice. } 

% Furthermore, imposing the 5D equation of motion for $h_{MN}$ constrains five more degrees of freedom, that we choose to be $\nabla^\mu h_{\mu\nu}$ and $h_\mu^\mu$.  

Linearizing the Einstein equation around this background 
  yields the equation of motion  for $h_{MN}$, 
\begin{equation}
   ^{(5)}{\cal D}^{MN}_{AB} h_{MN} \equiv   \,^{(5)}{\cal D}\,h_{AB} =0\,, \quad\quad\quad ^{(5)}{\cal D}^{MN}_{AB} = \delta_{A}^{M}\delta_{B}^{N}\square  + 2R^{~~M~~N}_{A~~B} \,. 
    \label{Lichnerowicz_Equation_Flat}
\end{equation}
This is sometimes called the Lichnerowicz equation.

The equation of motion implies  five relations between the remaining degrees of freedom. One can choose to eliminate $h_\mu^\mu$ and $h_{\mu y}$ in favor of the other components.  Finally, requiring the solutions to be well-behaved at both the horizon and at large $\rho$ forces the pieces  $h_{yy}$ to vanish for any unstable perturbation \cite{Gregory:2011GLInstability,Gregory_1994}. This implies that $h^\mu_\mu=0$, hence the only remaining component is the transverse-traceless component of $h_{\mu\nu}$.

We further decompose $h_{\mu\nu}$ in terms of the symmetries of the background metric. Considering translation invariance in both time and in the extra dimension, and $SO(3)$ of the   4D slices,  we can  decompose the fluctuation as 
\begin{equation}
    h_{\mu\nu} = \int_{-\infty}^{\infty} \frac{d\omega  dm}{(2\pi)^2} ~e^{i\omega t}e^{i m y} \ H^{\omega,m}_{\mu\nu}(\rho)\,.
    \label{perturb_GL_flat}
\end{equation}
Substituting  into \eqref{Lichnerowicz_Equation_Flat}  yields the final perturbation equation of motion
\begin{equation}
    \left(\,^{(4)}{\cal D} + m^2 \right) H^{\omega,m}_{\mu\nu} = 0\,,\quad\quad m\in \mathbb{R} \,. 
    \label{massive_lichnerowicz}
\end{equation}
The $\omega$ dependence is hidden inside the $\,^{(4)}{\cal D}^{\mu\nu}_{\rho\sigma}  $ operator. 

To establish instability of the black string, one searches for an unstable mode in \eqref{massive_lichnerowicz}, characterized by an imaginary value for $\omega$,  $\omega=-i\Omega$, with $\Omega>0$.  
 Numerical analyses performed by Refs.~\cite{GregoryLaflamme1993,Gregory_1994} revealed that  \eqref{massive_lichnerowicz} admits modes with $\Omega>0$ in a finite range of~$m$, 
 \begin{equation}
    0<|m|<\frac{k_{\nGL}}{\rho_{h}}\,,
\label{instabil.range}
\end{equation}
where  $m_{\nGL}\ = \frac{k_{GL}}{\rho_{h}} $ is the threshold mass for GL instability, $\rho_{h}$ is the black string radius and $k_{\nGL}\simeq 0.9$ for the five-dimensional case~\cite{Gregory_1994,Gregory:2011GLInstability}.\,\footnote{The $m=0$ mode is pure gauge in  full flat space~\cite{GregoryLaflamme1993}. In the presence of a brane,  it instead remains as the 4D graviton experienced by a brane-localized observer, which experiences the black string as a normal Schwarzchild black hole --- that is GL-stable. Therefore, in either cases, the $m=0$ mode does not cause GL instability. The solutions with  $m=0$ and $\omega=0$ are instead relevant to describe the static deformability of the black string, see e.g.~\cite{Hui:2020xxx,Barbosa:2025uau}. }

Finally, observe that, since $m\in \mathbb{R}$, for any $\rho_h$ there exists a value of $m$ in the interval \eqref{instabil.range}. Therefore black strings of any radius are unstable in flat space.

\section{Spacetime Fluctuations}
\label{eq:fluctuations}

To analyze the GL stability of the black string in ${\cal M}_\nu$, we first have to study the fluctuations of this spacetime. 

\subsection{General Equation of Motion and  Gauge Fixing }

The metric fluctuations take the general form ${g}_{MN} = \gamma^{\rm BS}_{MN} + \tilde{h}_{MN}$. Here $g_{MN}^{\rm BS}$ is the black string metric of \eqref{Black_String_Line_Element}. 
  Working in conformal coordinates, it is convenient to write the perturbation as 
\begin{equation}
  \gamma^{\rm BS}_{MN}+ \tilde{h}_{MN} = e^{-2A(z)} \left( \gamma^{\rm BS, flat}_{MN} + 2M_{5}^{-3/2} h_{MN} \right)\,. 
\end{equation}
Using this parametrization one can compute the fluctuations of the Ricci scalar using conformal rescaling $\gamma^{\rm BS}_{MN}= e^{-2A(z)}\gamma^{\rm BS, flat}_{MN}$, see~\cite{Fichet:2023dju}.

We now remove five degrees of freedom using the  gauge symmetry of $h_{MN}$, but with a different choice as in the flat space case of section \ref{se:GL_review}.     
Here we use the traceless-unitary gauge $h_{\mu 5}=0=h_\mu^\mu$,  as in \cite{Fichet:2023dju}. In this gauge, $h_{\mu\nu}$ is traceless but not transverse.

Then the quadratic action for $h_{\mu\nu}$ \cite{Fichet:2023dju} is simply 
\begin{equation}
\begin{split}
&S_{\rm quad} = \frac{M^{3}_{5}}{2}\int d^{4}x dz\sqrt{g} \ e^{-3A(z)}\left(\frac{1}{4}\nabla_{\rho}h^{\nu\sigma}\nabla^{\rho}h_{\nu\sigma} -\frac{1}{2}\nabla_{\mu}h_{\nu}^{\rho}\nabla_{\rho}h^{\mu\nu} - \frac{1}{4}\partial_{z}h_{\mu\nu}\partial_{z}h^{\mu\nu} \right)\,.
\label{onshellActionZcoord}
\end{split}
\end{equation}
Varying this action produces  the equations of motion  for the perturbation,
\begin{equation}
    e^{3A(z)}\partial_{z}(e^{-3A(z)}\partial_{z}h_{\mu\nu}(x^{\mu},z)) - \ ^{(4)}{\cal D}\,h_{\mu\nu}(x^{\mu},z) = 0\,,
    \label{pertEqMnun}
\end{equation}
where $^{(4)}{\cal D}_{\mu\nu}^{\rho\sigma}$ is the flat-space four-dimensional Lichnerowicz operator.
Let us notice the important fact that,  in any spacetime, since $^{(4)}{\cal D}\,h_{\mu \nu}(x^{\mu},z)$ is nothing but the linearization of the 4D Einstein tensor $^{(4)} G_{\mu\nu}$, the Bianchi identity implies
\be
\nabla^\mu \left[^{(4)} {\cal D}\,h_{\mu \nu }   \right] =0\,.
\label{eq:Bianchi}
\ee

We decompose the fluctuation as
\be
h_{\mu\nu}(x^\mu,z)=\sum_{m\in \Lambda}  \Psi_m(z) H^m_{\mu\nu}(x^\mu)\,, \label{eq:hmunu_decomposition}
\ee
where the profiles  along the fifth dimension ${\Psi}_{m}(z)$ form a complete basis and $H^{m}_{\mu\nu}(x^{\mu})$  are the corresponding 4D graviton modes. These are  gravitons with mass $m$,   satisfying the equation of motion
 $(^{(4)} {\cal D}  +m^{2}) H^{m}_{\mu\nu}(x^\mu)=0$. 
As a result, the profiles in the extra dimension obey
\begin{equation}
    \partial_{z}\left( e^{-3A(z)}\partial_{z}{\Psi_m}(z) \right) + e^{-3A(z)
    }m^{2}{\Psi}_{m}(z) = 0\,.
    \label{eq:graviton_EOM}
\end{equation}

The sum in \eqref{eq:hmunu_decomposition} is over the spectrum $\Lambda$, which can have both discrete and continuous components. 
\footnote{The spectrum is symmetric under $m\leftrightarrow -m$. 
One may define $\Lambda$ to cover only positive values of $m$, upon specifying that the $m\neq 0$ elements have multiplicity two while $m=0$ has multiplicity one. Here we define $\Lambda$ to cover both positive and negative values.  }

We can see that taking the divergence of  $^{(4)} {\cal D} H^{m}_{\mu\nu}(x^\mu) =-m^{2} H^{m}_{\mu\nu}(x^\mu)$  and using the identity \eqref{eq:Bianchi} implies $\nabla^\mu H^{m}_{\mu\nu} =0 $ for any mode with $m\neq 0$. In other words, the massive graviton modes are automatically transverse. Since they are also traceless, the  massive gravitons have been reduced to their five physical polarizations. 
For $m=0$, a residual gauge symmetry remains, that can be used to make the traceless massless graviton transverse.\,\footnote{At the level of the functional integrals, this is the gauge redundancy coming from the integration over the brane degrees of freedom, that implements  the Neumann boundary condition of $h_{\mu\nu}$.   }

Our gauge choice has the advantage of isolating the relevant transverse-traceless $h_{\mu\nu}$ component. The remaining $h_{55}$ degree of freedom mixes with the fluctuations of the dilaton field. A careful analysis of their equations of motion reveals that a single fluctuation propagates, see \cite{Fichet:2023dju} for details. 
 As shown in \cite{Gregory_1994},   scalar fluctuations do not cause GL instability.

\subsection{Fluctuations of  the ${\cal M}_\nu$ Spacetime with No Brane}

Our focus is on the fluctuations of the tensor modes. 
First consider the ${\cal M}_\nu$ space with no brane. 
The solutions to the equation of motion \eqref{eq:graviton_EOM} can be chosen as
\begin{align}
\Psi^m_{\pm}(z)&=z^\alpha J_{\pm\alpha}(mz)\,\quad\quad 
\alpha= \frac{4-\nu^2}{2(1-\nu^2)}\,, \quad\quad {\rm if} \quad\quad\nu\neq 1 \,, \label{eq:PsiJ}\\
\Psi^m_{\pm}(z)&=e^{\left({-\frac{3}{2}\eta\pm i \gamma_m }\right)z}\,\quad\quad 
\gamma_m= \sqrt{ m^2- \frac{9}{4}\eta^2 } \,, \quad\quad {\rm if} \quad\quad\nu=  1 \,.  
\end{align}

Regarding the $\nu\neq 1$ case, we have $\alpha<0$ for $1<\nu<2$, and $\alpha\geq2$ for $0\leq\nu<1$.
The solutions with Bessel order higher than $-1$ are square-integrable over the spacetime. This implies that $\Psi_+$ is integrable for $0\leq\nu<1$ and $\sqrt{2}<\nu<2  $, while $\Psi_-$ is integrable for $1<\nu<2 $.  The integrable solutions form a continuous orthogonal basis with 
\be
\int_{0}^\infty dz\,w(z)  \Psi^m_{\pm}(z)\Psi^{m'}_{\pm}(z) \propto \delta(m-m') \,,
\ee
with $ w(z)=z^{1-2\alpha}$.   The corresponding spectrum is
\be
\Lambda[{\cal M}_{\nu\neq 1}]=\mathbb{R}^*\,.
\ee
 The solutions with $m=0$ are $\Psi^0= \{cst,z^{2\alpha}\}$,  and thus do not produce normalizable modes. 

In the $\nu=1$ case, we have that both $\Psi_+^m$ and  $\Psi_-^m$ are square-integrable over the full domain $z\in \mathbb{R}$, and with $w(z)=e^{3\eta z}$.\,\footnote{In the LD case with no brane,  the warp factor is taken to be simply {$A(z)=- \eta z$}, as in \cite{Megias:2021mgj,Fichet:2023xbu}.}  The corresponding spectrum  is 
\be
\Lambda[{\cal M}_{\nu = 1}]=  \textstyle \left( -\infty ,-\frac{3}{2}\eta \right] \cup \left[ \frac{3}{2}\eta,\infty \right)\,.
\ee
There is thus a \textit{mass gap} in the fluctuations of linear dilaton spacetime, that is absent for ${\cal M}_{\nu\neq 1}$. The $m=0$ solutions are not normalizable.

\subsection{Fluctuations of  the ${\cal M}_\nu$ Spacetime with a Brane }

Previous results are for the ${\cal M}_\nu$ spacetime {in the absence of any brane}. We now consider the spacetimes truncated by the brane. 
Varying the action and taking into account the GHY term, the graviton field satisfies a Neumann boundary condition on the brane
\be
\partial_{z}\Psi^m|_{z=z_{b}}=0\,. \label{eq:BC_grav}
\ee

\subsubsection{Case $\nu\neq 1$ }

\subsubsection*{The $z\in (0,z_b]$ region }

We have seen that, for  $\nu<1$ and $1<\nu\leq \sqrt{2}$,  a single solution is normalizable. These are respectively $\Psi_+$ and $\Psi_-$. If $\nu>\sqrt{2}$, both $\Psi_\pm$ are normalizable,
however requiring regularity at the $z=0$ boundary  removes $\Psi_+$ anyways.

The brane boundary condition \eqref{eq:BC_grav} applied to the remaining solution is  
\begin{align}
&\partial_{z}\Psi_+^m|_{z=z_{b}}=0\,,\quad\quad  {\rm if} \quad\quad\nu< 1\,, \nn \\
&\partial_{z}\Psi_-^m|_{z=z_{b}}=0\,,\quad\quad  {\rm if} \quad\quad\nu> 1\,. 
\end{align}
 This condition constrains $m$ to take discrete values. We find 
 \begin{align}
 m_n\big|_{\nu<1} =\frac{j^{(n)}_{\alpha-1}}{z_b} \,,\quad\quad 
  m_n\big|_{\nu>1}=\frac{j^{(n)}_{1-\alpha}}{z_b} \,,
 \end{align}
 with $j_{\alpha}^{(n)}$ the $n$-th zero of $J_{\alpha}$. 

In the  $m=0$ case, the $z^{2\alpha}$ mode is removed by the boundary condition and the $\Psi^0=cst$ mode becomes normalizable only for $\nu>1$, for which $\int dz w(z)|\Psi^0(z)|^2 $ is finite. 

\vspace{0.5cm}
\subsubsection*{The $z\in [z_b,\infty)$ region }

In this region, both $\Psi_\pm$ solutions are simultaneously normalizable because the singularity is outside the region. As a result, the boundary condition fixes the relative constant between  $\Psi_+$ and $\Psi_-$, but does not constrain the spectrum. In the  $m=0$ case, the $\Psi^0=cst$ mode becomes normalizable only for $\nu<1$. 
We conclude that the spectrum is $\mathbb{R}$ for $\nu<1$ and $\mathbb{R}^*$ for $\nu>1$. 

\subsubsection{Case $\nu =  1$ }

In the presence of the brane, the mass gap is given by $\mLD\equiv \frac{3\eta r_b}{2L}$,  the profiles are  $e^{\left( -\mLD\pm i \gamma_m \right) z} $, and $w(z)=e^{2\mLD z}$. Since both $\Psi_\pm$ solutions are normalizable over the whole space, the boundary condition on the brane constrains only their linear combination and the spectrum is not affected.

On the other hand, in the  $m=0$ case, the $\Psi^0 = e^{-2\mLD z} $ mode is removed by the boundary condition and the constant mode  becomes normalizable on $(-\infty, z_b)$ since $\int dz w(z)|\Psi^0(z)|^2 $ is finite there. The spectrum is thus $\left( -\infty ,-\mLD \right] \cup \{0\} \cup \left[ \mLD,\infty \right)$ on $(-\infty, z_b]$, and $\left( -\infty ,-\mLD \right] \cup  \left[ \mLD,\infty \right)$ on $[z_b,\infty)$.

\subsection{Spectrum Summary \label{se:spectrum}}

We summarize  the spectrum of gravitational fluctuations  in Table\,\ref{tab:spectrum}.   We 
 use the definitions \eqref{eq:Mpmdefconf} to state coordinate-independent results.

\begin{table}[h]
    \centering
    \begin{tabular}{|c|c|c|c|}
    \cline{2-4}
        \multicolumn{1}{c|}{} & $\nu<1$ & $\nu=1$ & $\nu>1$  \\
         \hline 
        $\Lambda[{\cal M}^-_\nu ] $    & 
          $\mathbb{R}$  &  $\left( -\infty ,-\frac{3 \eta r_b}{2L} \right] \cup \{0\} \cup  \left[ \frac{3 \eta r_b}{2L},\infty \right)$   &  $\left\{0, \frac{j^{(n)}_{1-\alpha}}{z_b}\right\}$ \\
        \hline
            $ \Lambda[{\cal M}^+_\nu ] $     &    $\left\{\frac{j^{(n)}_{\alpha-1}}{z_b}\right\}$   & $\left( -\infty ,-\frac{3 \eta r_b}{2L} \right] \cup \left[ \frac{3 \eta r_b}{2L},\infty \right)$     &   $\mathbb{R}^*$   \\ 
        \hline
    \end{tabular}
    \caption{ Spectrum  of gravitational fluctuations in the ${\cal M}^\pm_\nu$  spacetimes.   }
    \label{tab:spectrum}
\end{table}

 We can see that the massive spectrum is discrete for ${\cal M}^+_{\nu<1}$ and ${\cal M}^-_{\nu>1}$. These are the regions that have the $z=0$ boundary --- which is either conformal or regular, as established in section \ref{se:boundaries}. As can be seen from \eqref{eq:PsiJ}, the modes tend to be repelled from the boundary. They are thus effectively  confined between the boundary and the brane, and the resulting spectrum is discrete.

The logics is different for the massless mode. We can see that the zero mode is always in the ${\cal M}_\nu^-$ spacetime, which is the side that  contains the curvature singularity. 
This is best understood by absorbing the weight function into the field, $\tilde \Psi^0=\sqrt{w}\Psi^0$. What happens is that  $\tilde \Psi^0$ tends to be repelled from the singularity --- which coincides with the boundary only for $\nu>1$. The massless mode becomes normalizable when  it is trapped between the brane and the singularity,   which occurs in ${\cal M}_\nu^-$.

While the presence of the graviton zero mode is irrelevant for the black string stability, it would be essential in the scope of  building a ${\cal M}_\nu$ -- braneworld model  featuring low-energy gravity \cite{Fichet:2023dju}. Although  the decoupling of gravity at low-energy might seem surprising, it does happen in well-defined theories such as Little String Theory \cite{Aharony:1999ks,Kutasov:2001uf}, which  is reminiscent of the $\nu=1$ spacetime \cite{Aharony:1998ub}.

\section{Stability of the Black String}
\label{se:stability_results}

\subsection{Results}

Our analysis of the metric fluctuations has reduced the problem to 
\be
\left( \,^{(4)} {\cal D} +m^{2} \right) H^{m}_{\mu\nu}(x^\mu) =0 \,,\quad\quad m\in \Lambda\left[{\cal M}_\nu\right] \,.
\label{eq:EOM_4Dmodes}
\ee
The nontrivial information in \eqref{eq:EOM_4Dmodes} is encoded into the  spectrum in spacetime ${\cal M}_\nu$, denoted by $\Lambda[{\cal M}_\nu]$.

The standard GL analysis reviewed in section \ref{se:GL_review} directly applies to \eqref{eq:EOM_4Dmodes}, the outcome being that the black string is unstable if there is a mode $m$ in the GL interval 
\be \Lambda_{\rm GL}\equiv \left(-\frac{k_{\rm GL}}{\rho_h}, 0 \right) \cup \left(0,\frac{k_{\rm GL}}{\rho_h} \right)\,. \ee
Therefore, the problem boils down to studying the intersection of $\Lambda\left[{\cal M}_\nu\right]$  with $\Lambda_{\rm GL}$. For a given black string radius $\rho_h$, we have that  
\be
\Lambda\left[{\cal M}_\nu\right] \cap \Lambda_{\rm GL} \neq \emptyset \quad \Longleftrightarrow \quad  {\rm Classically~unstable~black~string }\,.  
\ee
Conversely, $\Lambda\left[{\cal M}_\nu\right] \cap \Lambda_{\rm GL} = \emptyset$ implies that there is no GL instability.  

Since the  range $\Lambda_{\rm GL}$ depends on $\rho_h$,  the essential point is whether or not the spectrum has a mass gap. If the spectrum has no mass gap, i.e.~extends down to zero, for any black string radius $\rho_h$ there are modes that make the black string unstable. 

In constrast, if the spectrum has a mass gap $m_g>0$, then there is no mode availaible if $\frac{k_{\rm GL}}{\rho_h}<m_g$,  i.e. one has $\Lambda\left[{\cal M}_\nu\right] \cap \Lambda_{\rm GL} = \emptyset$. In other words, if the black string radius is larger than the critical radius $\rho_h^c\equiv \frac{k_{\rm GL}}{m_g} $, then it is stable.

The stability criterion applied to our results of Tab.\,\ref{tab:spectrum}, combined with the definitions \eqref{eq:Mpmdefconf}, provides the following coordinate-independent conclusion:  \\
\begin{table}[h]
    \centering
    \begin{tabular}{|c|c|c|c|}
    \cline{2-4}
        \multicolumn{1}{c|}{} & $\nu<1$ & $\nu=1$ & $\nu>1$  \\
         \hline 
        ${\cal M}^-_\nu$     &  unstable  &    stable/unstable  & stable/unstable  \\
        \hline
            ${\cal M}^+_\nu$     &   stable/unstable  &  stable/unstable    &  unstable \\
    \hline 
    \end{tabular}
    \caption{ Classical stability of the black string  in the ${\cal M}_\nu^\pm$ spacetimes. In the stable/unstable cases, black strings with radius $\rho_h>\rho_h^c$ are stable, and are unstable otherwise.   }
    \label{tab:stability}
\end{table}

\subsection{Discussion}

We see that the black string is unstable for ${\cal M}_{\nu<1}^-$ and  ${\cal M}_{\nu>1}^+$. These are the regions in which there is no timelike boundary, neither conformal nor regular. Conversely, the black string {with $\rho_h > \rho_h^c$} is stable for ${\cal M}_{\nu>1}^-$ and ${\cal M}_{\nu<1}^+$, for which the black string ends on the timelike boundary, either conformal or regular. 
Notice that the conformal coordinate $z$  renders manifest the connection between  the conformal boundary of ${\cal M}_{\nu<1}^+$ and the discrete spectrum.  This feature would remain obscured if one used the proper coordinate $r$. 

The black string horizon area is finite for ${\cal M^-_\nu}$ and is infinite for ${\cal M^+_\nu}$, for all $\nu$. In the ${\cal M}^-_{\nu>1}$ case, which has a regular boundary,  the black string  extends along a compact direction, therefore its stability is analogous to that of a black string wrapping on $S_1$ or stretched on a compact interval.  
 In our ${\cal M}_{\nu}$ background, the compact direction is a consequence of the  spacetime curvature itself.

In the ${\cal M}^+_{\nu<1}$ case, where the timelike boundary is conformal, there is no curvature singularity and the black string area is infinite. Nevertheless, perhaps surprisingly, the black string solution remains classically stable.
This is also true for the the LD spacetime ($\nu=1$), for which the conformal boundary is null (i.e. identical to the one of Minkowski spacetime). 
In LD$^+$, the black string has infinite area, there is no curvature singularity, and still, the black string is stable.

\subsubsection{On Thermodynamic Stability}

Our black string is thermodynamically unstable in any region and for any $\nu$. This can be easily seen because, for any slices with constant $r$, the string simply amounts to a Schwarzschild black hole --- which is thermodynamically unstable since it has negative specific heat. 

It has been conjectured that, in geometries with a non-compact translation symmetry, classical  and thermodynamic stabilities are related \cite{Gubser:2000ec,Gubser:2000mm}. Pieces of evidence with various levels of generality and rigor support this correlated-stability conjecture (CSC) see e.g. \cite{Reall_2001, Friess_2005,BRIHAYE2008264,Dhumuntarao:2022,Miyamoto_2008}.
Strictly speaking, such considerations do not apply to our spacetime, which does not have translational invariance for any $\nu$. It is nevertheless interesting to discuss the correlations (and lack thereof) between classical and thermodynamic stabilities that appear in our case.

From Table \ref{tab:stability}, we have that the black string is stable for ${\cal M}_{\nu>1}^-$ and ${\cal M}_{\nu<1}^+$, and is  unstable for ${\cal M}_{\nu<1}^-$ and  ${\cal M}_{\nu>1}^+$, in which cases  there is no  timelike boundary (neither conformal or regular). 
These facts suggest the absence of a timelike boundary as a relevant criterion for applicability of thermodynamic arguments.  
We notice however that in the critical  case of $\nu=1$, on both sides of the brane, the boundary is null, just like in flat space, and yet the black string is stable.

It  has been argued  that finite volume effects spoil thermodynamic arguments \cite{Gubser:2000mm}, in which case the CSC does not apply. 
Here we see that the  black string  area does not influence the classical stability: the black string has finite area in ${\cal M}_\nu^-$ and infinite area in ${\cal M}_\nu^+$, for all $\nu$, hence all combinations happen. We conclude that the (lack of) timelike boundary on which the string ends seems to be a more relevant quantity when  attempting  to drop the  translational symmetry condition from the CSC.

 \subsubsection{Holographic confinement}

 The ${\cal M}_\nu$ spacetimes have a holographic intepretation  in terms of an ---a priori unknown--- strongly-interacting theory coupled to 4D gravity \cite{Barbosa:2024pyn}.  The putative strongly-coupled theory can  exhibit a notion of confinement, that can be assessed using criteria such as the behavior of classical strings,  of entanglement entropy, and of the thermodynamical phase transitions.   Whether the ${\cal M}^-_\nu$  spacetime is confining  has  been analyzed in details in \cite{Barbosa:2024pyn}, where it was found that confinement occurs for $\nu \geq 1$. 
 
This is precisely the range for which stable black strings exist. 
Since the black string simply is a Schwarzschild black hole in the dual 4D theory, such a correlation suggests an interesting interplay between black holes and the confining sector. We will investigate this interplay in a future work.

\section{Conclusion \label{se:Conclu}}

Can spacetime curvature stabilize black strings? In this work, we have shown that it can.

Our framework is a  5D dilaton-gravity system stabilized by the presence of a flat brane. The resulting class of spacetimes is denoted  ${\cal M}_\nu$. The brane splits the spacetime into two inequivalent regions, ${\cal M}_\nu^\pm$. Depending on the bulk potential and on the region considered, this class of spacetimes may exhibit a regular boundary, a conformal boundary, and/or a curvature singularity. 
In a sense, the ${\cal M}_\nu^\pm$ spacetimes offer a variety of environments for the black string.

As an aside, we propose a new criterion to determine whether a curvature singularity is admissible. This criterion is based on spacetime fluctuations and does not require the construction of a planar black hole solution; it merely amounts to checking for the existence of a graviton zero mode. 

We  first show that the black string solution exists within the ${\cal M}_\nu^\pm$ spacetimes. We then systematically explore the (GL) classical stability of the black string in ${\cal M}_\nu^\pm$.

To this end, we compute the spectrum of fluctuations in the ${\cal M}_\nu^\pm$ spacetimes.  
We find that the spectrum is discrete whenever the geometry exhibits either a regular or a conformal boundary. Moreover, a zero mode is present when the spacetime contains a (good) curvature singularity, because the zero mode is effectively repelled from the singular region.

Combining the spectrum computation with the standard GL result,  assessing classical stability simply amounts to determining whether the spectrum is gapped. We find that the  black string is unstable for ${\cal M}^+_{\nu>1}$ {and ${\cal M}^-_{\nu<1}$}, and is stable for sufficiently large radius  in all other cases, see Table \,\ref{tab:stability}.

In the ${\cal M}^-_{\nu>1}$ case, which has a regular boundary,  the boundary coincides with the singularity, and the  black string horizon area is finite. The black string therefore extends along a compact direction, and its stability is analogous to that of a black string wrapping on $S_1$ or stretched on a compact interval. Our ${\cal M}_{\nu}$ background illustrates that spacetime curvature alone is able to produce a compact direction.

In the ${\cal M}^+_{\nu<1}$ case, where the boundary is conformal, there is no curvature singularity and the black string area is infinite. Nevertheless, perhaps surprisingly, the black string solution remains classically stable.
Moreover, in the critical case of the linear dilaton spacetime ($\nu=1$), the black string is stable in both LD$^-$ and LD$^+$, even though in the latter the area is infinite and no curvature singularity is present.

The correlated-stability conjecture, which relates thermodynamic and classical stability,  does not apply to our black string as it lacks  translational invariance.  Still, if one wishes to generalize the CSC by  dropping the assumption of translational invariance, our results show that finite-volume effects are not the only obstruction. A more general requirement of no timelike boundary --- whether regular or conformal --- is necessary. A generalization of the CSC would also have to account for the linear dilaton case, for which there is no timelike boundary, black strings have either finite (LD$^-$) or infinite (LD$^+$) area, and yet are stable in either cases.

Finally, for a brane observer, the black string appears as a Schwarzschild black hole, while the bulk dynamics gives rise to a strongly-interacting sector due to the gauge–gravity correspondence. Remarkably, black string stability is precisely correlated with confinement in the strongly-coupled sector, see \cite{Barbosa:2024pyn}. We will explore this connection in future work.

\begin{acknowledgments}

EM would like to thank the ICTP South American Institute for Fundamental Research (SAIFR), S\~ao Paulo, Brazil, for hospitality and partial financial support during the final stages of this work. SF is supported by grant 2021/10128-0 of FAPESP. The works of SF, EM and MQ are supported by the ``Proyectos de Investigaci\'on Precompetitivos'' Program of the Plan Propio de Investigaci\'on of the University of Granada under grant PP2025PP-18. The work of MQ is also supported by the grant PID2023-146686NB-C31 funded by MICIU/AEI/10.13039/501100011033/ and by FEDER, EU. IFAE is partially funded by the CERCA program of the Generalitat de Catalunya. The work of GY is supported by Grant No. 001 of CAPES.

\end{acknowledgments}

\appendix

\bibliographystyle{JHEP}
\normalem
\bibliography{biblio}

\end{document}